\documentclass{article}
\usepackage{moriond}
\usepackage{rotating}
\usepackage{graphics}
\usepackage{epsfig}
\def\bm{\mathbf}
\def\Journal#1#2#3#4{{#1} {\bf #2}, #3 (#4)}
\begin{document}

\vspace*{4cm}
\title{Multi-frequency Wiener filtering of CMB data with polarization
  and estimation of cosmological parameters}

\author{ S. Prunet$^{1}$, S.K. Sethi$^{2}$, F.R. Bouchet$^{2}$ }

\address{${}^{1}$IAS, b\^at. 121, 91405 Orsay\\ 
${}^{2}$IAP, 98bis Boulevard Arago, 75014 Paris }

\maketitle\abstracts{We present a method of subtracting the foreground
contamination for the measurement of CMB polarization. We calculate the
resultant errors on CMB polarization and temperature-polarization
cross correlation power spectra for the high frequency instrument
(HFI) aboard Planck Surveyor, and estimate the corresponding errors
on cosmological parameters}

\section{Introduction}

The upcoming satellite CMB experiment Planck surveyor offers an
unprecedented opportunity to measure CMB polarization. 
A major hurdle in extracting the primary CMB signal from data, apart
from noise, is galactic and extragalactic foregrounds. However, as the
foregrounds differ from CMB in both  frequency dependence and spatial
distribution, one can hope to reduce their level in a multi-frequency 
CMB experiment. A multi-frequency Wiener filtering method to implement
this scheme  was developed \cite{bouchet,teg} 
 and applied to cleaning the  simulated CMB temperature
 map   by   Bouchet {\em et al.}
 \cite{bouchet}. 
 They showed  that the residual contamination after cleaning
the map is much smaller than the CMB primary signal, and therefore 
the foregrounds may not be a major obstacle in 
 the extraction of  CMB temperature
angular power spectrum.  However, as the 
CMB  polarization signal is expected to be one to two orders of
magnitudes below the temperature signal, it is likely to
be comparable to both  the experimental  noise and the level of
foregrounds. Sethi {\em et al.}  \cite{set}  modelled and
estimated the level of dust polarized
emission at high galactic latitudes. Dust polarization is expected to
be the dominant contaminant for  measuring CMB polarization using
Planck HFI.

In this paper, we extend the multi-frequency Wiener filtering method
to include the polarization and temperature-polarization
cross-correlation. The aim of this exercise is to quantify errors in
estimating various power spectra and consequently the errors in
cosmological parameters. Our results are relevant for Planck HFI.

\section{Multi-frequency Wiener filtering on CMB data}

Let us denote the observed data at different  frequencies as,
$y_{\nu}^i$, 
where $\nu$ indicates the frequency of the instrumental channel, and $i$ the
nature of the observed field (temperature $T$ and $E$-mode 
polarization\cite{sel}). $y_{\nu}^i$, for a given $i$,
 takes contributions from 
various galactic and extragalactic sources, apart from the primary CMB
signal and instrumental noise. 
 Let us call $x_p^j$ the contribution   of the field $j$ due to
process $p$, 
this is
the quantity we want to recover from the observational data $y_{\nu}^i$.
We assume that there is  a linear relation between them. In multipole
space, it can be expressed as:
\begin{equation}
y_{\nu}^i (l,m)=A_{\nu p}^{ij} (l,m) x_p^j (l,m) + b_{\nu}^i (l,m)
\label{eq:filmod}
\end{equation}
where $A_{\nu p}^{ij}$ is the instrument response kernel, and $b_{\nu}^i$ is
the detector noise level per pixel for the full mission time (all repeated
indices are summed over). 

The problem is now how to construct an optimal estimator $\hat x_p^j$. We
chose this estimator to be linear in the observed data:
\begin{equation}
\hat x_p^i = W_{p\nu}^{ij} y_{\nu}^j 
\label{eq:recon}
\end{equation}
The reconstruction error for a given field and 
 process  is given by $|(\hat x_p^i - x_p^i)^2| \equiv
(\varepsilon_p^i)^2$. 
$W_{p\nu}^{ij}$ is chosen so as to make 
 the variance of the reconstruction error minimal. 
 The derivatives of the error with respect
to the filters coefficients $W_{p\mu}^{ic}$ should then be zero. This
 condition can be  expressed as:
\begin{equation}
A_{\mu p'}^{ck}W_{p\nu '}^{il}A_{\nu 'p''}^{lm} \langle x_{p'}^k
x_{p''}^m \rangle
+ W_{p\nu '}^{ib} \langle b_{\mu}^c b_{\nu '}^b \rangle =
 A_{\mu p'}^{ck} \langle x_{p'}^k
x_p^i \rangle.
\label{eq:wien}
\end{equation}

\subsection{Implementation of foregrounds removal}

Eq.~(\ref{eq:wien}) is valid for the general case in which various
processes, fields, and corresponding instrumental noises
 could be correlated.  
We consider here only uncorrelated processes and noises between
different fields and channels. We allow for the correlation between the two
fields $T$ and $E$. 
With these conditions, Eq.~(\ref{eq:wien}) can be written as  a
 system of four matrix equations which
 can be solved by substitution. 

Bouchet {\em et al.} \cite{bouchet}  defined a quantity called
the 'quality factor' to understand the
quality of extraction of the signal corresponding to a given process. 
A  straightforward generalisation of the  quality
factor, valid for multiple fields, can be written as:
\begin{equation}
Q_{pp'}^{ij}=\frac{ \langle \hat x_p^i \hat x_{p'}^j \rangle}{ \langle
x_p^i x_{p'}^j \rangle} = 
W_{p\nu}^{ik} A_{\nu p''}^{kl} \langle x_{p''}^l x_{p'}^j \rangle
\end{equation}
where we have used  Eq.~(\ref{eq:wien}) to write the second equality.
 ${\bm Q}^{11}$ and ${\bm Q}^{22}$ can readily be interpreted as
the quality of the reconstruction of temperature and
polarization maps, respectively.  It should be noted that in the
presence of cross-correlations, the quality factor of either field
is better than the case without cross-correlations. Though the
reconstruction of temperature maps is only slightly changed by
cross-correlation term, the quality of polarization reconstruction gets
 a big boost
from the presence of temperature-polarization cross-correlation.
 However the meaning of 
  term ${\bm Q}^{21}$ (and ${\bm Q}^{12}$) is not apparent. 
Much of the contribution to ${\bm Q}^{12}$ comes from the term
with ${\bm W}^{11}$, and therefore it is very close to the quality
factor for the extraction of temperature and is nearly independent of 
the polarization noise.  It is not surprising as it
merely tells us that to optimally
reconstruct the cross-correlation one needs to throw out the noisy
data, i.e. the polarization. However, the quantity of interest for us
is the error in the extraction of the power spectrum of
cross-correlation, which should not  be confused with ${\bm Q}^{12}$. 
To get a real idea of the error bars of the different spectra, we must
define estimators of those power spectra from the filtered data, and
compute their covariance.

\subsection{Unbiased estimators of power spectra}

$\hat  x_p^i$ (Eq.~(\ref{eq:recon})) is the data obtained after 
performing  Wiener filtering on  the  multi-frequency maps. Our aim in
this section is to use this data to write an unbiased estimator of the
true power spectrum  $x_p^i$. 
The average  power
spectrum of $\hat  x_p^i$ can be written as:
\begin{equation}
\langle \hat  x_p^i \hat  x_p^i \rangle   = W_{p\nu}^{ij}
 W_{p\nu '}^{il} \left [A_{\nu p'}^{jk} A_{\nu ' p''}^{lm} \langle
 x_p^i  x_p^i \rangle  + \langle b_\nu^j b_{\nu '}^l \rangle \right ]
\end{equation}

This can be expanded to give:
\begin{equation}
\langle \hat x_p^i \hat x_p^j \rangle  =  (Z_p^{ij} C_p^{ij} + b_p^{ij}) = 
 Q_p^{ij} C_p^{ij}
\end{equation}
Here $\{i,j\}$ stand for $T$ and $E$. 
 $b_p^{ij}$ are generalized versions of the instrumental 
noise, containing contributions of the noise from  different channels
 as well as some power leakage of the other processes and the other fields.
We then define unbiased estimators of the spectra:
\begin{equation}
\hat C_p^{ij}  =  \frac{1}{Z_p^{ij}} \left( 
\frac{1}{2\ell +1}\sum_{m} \| \hat x_p^{i}(m) \hat x_p^{j}(m)\|
-b_p^{ij} \right) 
\end{equation}
and compute their covariances:
\begin{eqnarray}
Cov(\hat C_p^{TT}) & = & \frac{2}{2\ell+1}\left(
C_p^{TT} Q_p^{11}/Z_p^{11} \right)^2 \\
Cov(\hat C_p^{EE}) & = & \frac{2}{2\ell+1}\left(
C_p^{EE} Q_p^{22}/Z_p^{22} \right)^2 \\
Cov(\hat C_p^{TE}) & = & \frac{1}{2\ell+1}\frac{\left(
(Q_p^{12})^2(C_p^{TE})^2 + Q_p^{11}Q_p^{22}C_p^{TT} C_p^{TE}
\right)}{(Z_p^{12})^2} 
\end{eqnarray}
It has been assumed that both CMB and foregrounds are Gaussian fields
in the computation of covariances. In Fig.~1 we show the E-mode
polarization and $ET$
 cross-correlation power spectra for sCDM model and the expected
 $1\sigma$ errors on their measurement using the specifications of
 Planck HFI. Apart from polarized dust, an additional contribution to
 polarized foregrounds from $40\%$ polarized synchrotron emission  has been
 assumed. As the synchrotron foreground is 
 subdominant at HFI frequencies, it doesn't affect our
 results. However, it is expected to be the major foreground for MAP
 and Planck low frequency instrument (LFI).

\begin{figure}
\centering\epsfig{figure=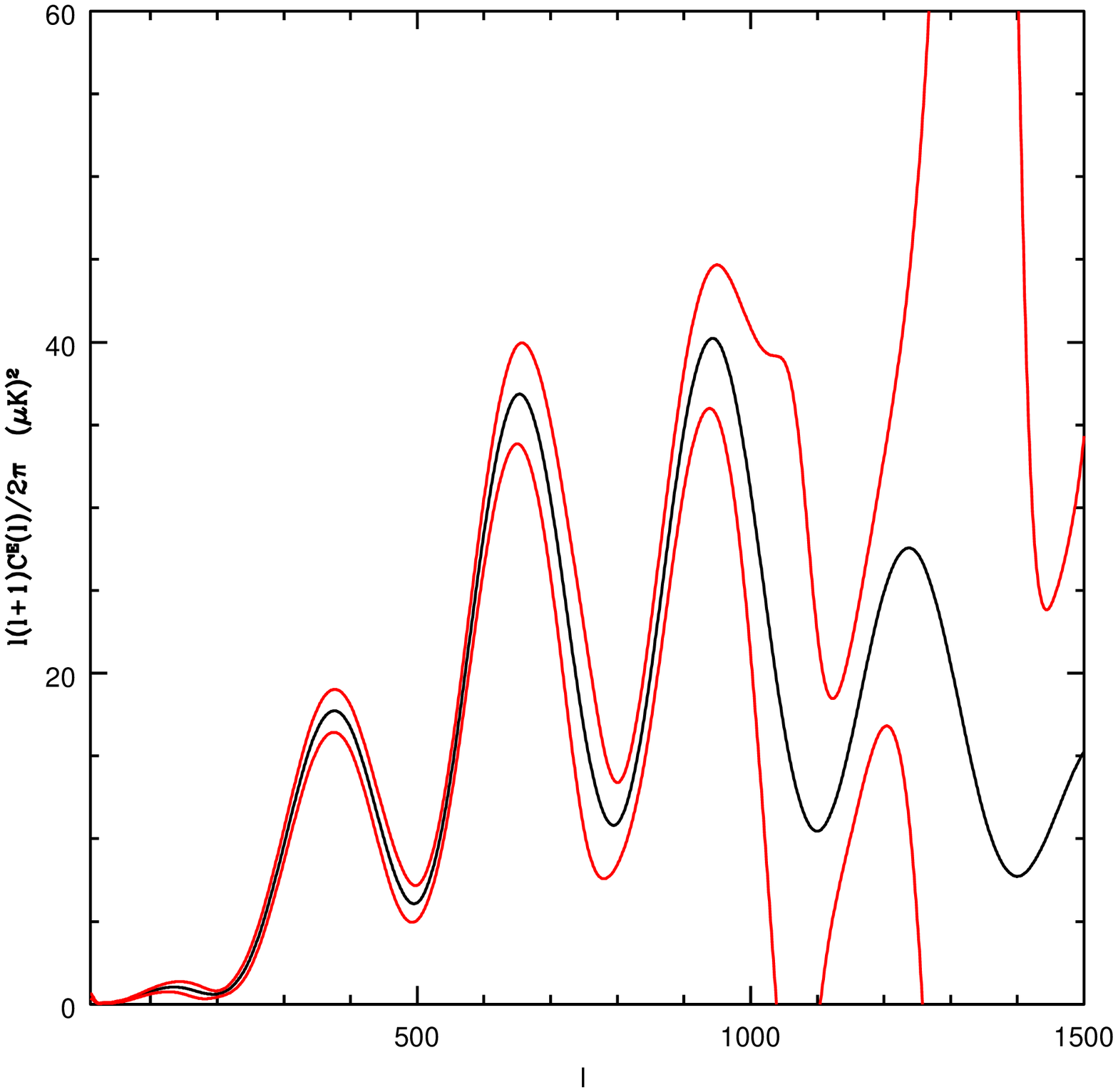,width=6cm,height=5cm}
\centering\epsfig{figure=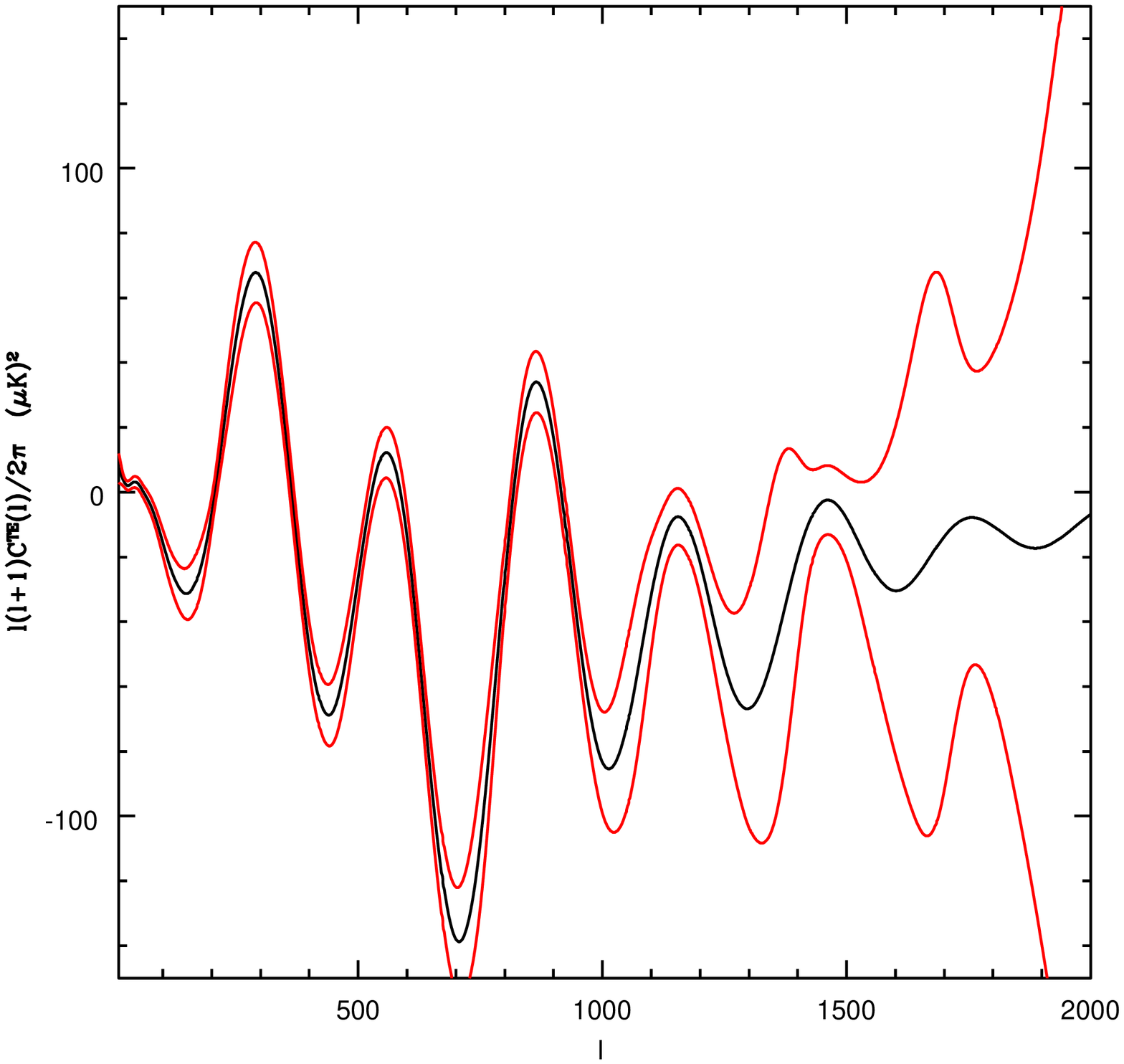,width=6cm,height=5cm}
\caption{Power spectra and $1\sigma$ errors for the E-mode polarization
and the temperature-polarization cross-correlation respectively. }
\end{figure}
\section{Errors on cosmological parameters}

The unbiased estimators defined above and their covariances
can be used to estimate the precision we expect on the 
measurement of cosmological  parameters once the foregrounds
are filtered out  for any desired experiment.

\subsection{Fisher Matrix}
The Fisher matrix is defined as an average value of the second
derivatives of the logarithm of the Likelihood function 
with respect to the cosmological
parameters, taken at the maximum likelihood  point:
\begin{equation}
F_{ij}=\left\langle\frac{\partial^2\mathcal{L}}{\partial \theta_i
\partial\theta_j}\right\rangle_{\Theta=\Theta_0}
\end{equation}
One can take this matrix as an estimate of the covariance
matrix of the cosmological parameters.
Following Zaldarriaga {\em et al.} \cite{zal} we generalize the
Fisher matrix approach to the polarized case. It can be written
as:
\begin{equation}
F_{ij}=\sum_{\ell=2}^{+\infty}\sum_{X,Y}
\frac{\partial C_{\ell}^X}{\partial \theta_i}
\rm{Cov}^{-1}\left(C_{\ell}^X,C_{\ell}^Y\right)
\frac{\partial C_{\ell}^Y}{\partial \theta_j}
\end{equation}
where $\rm{Cov}(C_{\ell}^X,C_{\ell}^Y)$ stands for
the covariance matrix of the power spectra estimators,
and $\{X,Y\} \in \{T,E,TE\}$.
We will then use the covariances computed for the 
unbiased estimators defined in the previous section
to take into account the foregrounds removal in 
the error bars computation.

With the present specifications of
Planck HFI, our  results are shown in Table~1 for a variant of CDM
model. The best channel case corresponds to $\nu = 143 \, \rm GHz$ channel
with no foregrounds  and the worst case corresponds to the same
channel with foregrounds added as noise. As is clearly seen,
significant improvement on parameter estimation can be achieved if
the foregrounds are removed using the Wiener filtering technique
discussed in previous sections, except for the optical depth to the
last scattering surface $\tau$. An accurate determination of such
small values of $\tau$
depends  on  the level
of CMB polarization  at $\ell \le 10$,
 which in our case is  dominated
 by polarized foregrounds which cannot be easily removed. And
therefore,  a  careful
analysis of polarized foregrounds at large scales is necessary to
determine such a small value of $\tau$. 
The different power spectra were computed with the CMB Boltzmann
code CMBFAST \cite{cmbfast}.
\begin{table}
\caption{Errors on cosmological parameters.
$\Delta\Omega_\Lambda$ is given in absolute value.}
\scalebox{0.95}{\begin{tabular}{|l|c|c|c|c|c|c|c|c|}
\hline
\hline
Parameters & $C_2$ & $h$ & $\Omega_b$ & $\Omega_{\Lambda}$ & $\tau$ &
$n_s$ & $n_t$ & $T/S$\\
\hline
Model & $796 (\mu K)^2$ & $0.5$ & $0.05$ & $0.0$ & $0.05$ & $0.99$ &
$0.01$ & $0.1$  \\
\hline
Worst case & {$3.72$ \%} & {$1.11$ \%} & {$1.93$ \%} & $0.032$ &
{$28.3$ \%} & {$0.40$ \%} & {$538$ \%} & {$78.0$ \%}\\
\hline
Best channel & {$2.5$ \%} & {$1.07$ \%} & {$1.87$ \%} & $0.031$ &
{$2.74$ \%} & {$0.38$ \%} & {$453$ \%} & {$67.6$ \%}\\
\hline
Wiener & {$1.98$ \%} & {$0.23$ \%} & {$0.34$ \%} & $6.6\,10^{-3}$ &
{$19.7$  \%} & {$0.10$ \%} & {$242$\,\%} & {$27.0$ \%} \\ 
\hline
\hline
\end{tabular}}
\end{table}


\section*{References}
\begin{thebibliography}{99}
\bibitem{bouchet} F. R.  Bouchet  {\em et al. },  Space Science Rev. 74,
37(1995). 

\bibitem{teg} M. Tegmark and G.   Efstathiou, \Journal{ApJ}{281}{1297}{1996}.

\bibitem{set} S. K. Sethi, S. Prunet,  and F. R. Bouchet,  this volume.

\bibitem{zal} M. Zaldarriaga, D. N. Spergel, and  U. Seljak,
\Journal{ApJ}{488}{1}{1997}. 

\bibitem{sel}  U. Seljak, \Journal{ApJ}{482}{6}{1997}. 

\bibitem{cmbfast} U. Seljak and M.  Zaldarriaga,\Journal{ApJ}{469}{437}{1996}


\end{thebibliography}
\end{document}